\documentclass[
aps, 
prb, 
10pt,  
amssymb, 
amsmath, 
superscriptaddress, 
tightenlines, 
twocolumn, 
notitlepage, 
] {revtex4-2}
\usepackage{graphicx}
\usepackage[colorlinks, linkcolor = blue, citecolor = blue, urlcolor=blue, pdfborder={0 0 0 [0 0]},bookmarks=false]{hyperref}
\usepackage{bm}
\usepackage{physics}
\usepackage{amssymb}
\usepackage{amsmath}
\usepackage[dvipsnames]{xcolor}
\usepackage{soul}
\usepackage{lipsum}

\newcommand{\be}{\begin{equation}}
\newcommand{\ee}{\end{equation}}
\newcommand{\lp}{\left (}
\newcommand{\rp}{\right )}

\begin{abstract} 
Itinerant spin polaron and metallic ferromagnetism are theoretically predicted in the Mott insulator in semiconductor moir\'e superlattices doped below and above half filling of the narrow moir\'e band, respectively. The existence of spin polaron can be directly identified from the kink in the dependence of the charge gap on the magnetic field.
\end{abstract}

\begin{document}

\title{Itinerant spin polaron and metallic ferromagnetism in \\ semiconductor moir\'e superlattices}
\author{Margarita Davydova}
\affiliation{Department of Physics, Massachusetts Institute of Technology, Cambridge, MA 02139, USA }
\author{Yang Zhang}
\affiliation{Department of Physics, Massachusetts Institute of Technology, Cambridge, MA 02139, USA }
\author{Liang Fu}
\affiliation{Department of Physics, Massachusetts Institute of Technology, Cambridge, MA 02139, USA }

\maketitle

Recent experiments have discovered a plethora of novel electronic phases in  transition metal dichalcogenides (TMDs) heterostructures,  including Mott-Hubbard and charge transfer insulators~\cite{PhysRevLett.121.026402,PhysRevB.102.201115,regan2020mott,tang2019wse2,li2021continuous,https://doi.org/10.48550/arxiv.2202.02055}, generalized Wigner crystals~\cite{regan2020mott,xu2020correlated,zhou2021bilayer,jin2021stripe,li2021imaging,huang2021correlated}, the quantum anomalous Hall state~\cite{li2021quantum} and light-induced ferromagnetism~\cite{wang2022light}. 
These remarkably rich phenomena result from strong interaction effects in narrow moir\'e bands, which generally appear in TMD heterostructures with large moir\'e wavelengths.  
Take the example of WSe$_2$/WS$_2$: the lattice corrugation introduced by the moir\'e structure produces a periodic spatial variation of the valence band edge, which acts as a superlattice potential for charge carriers in WSe$_2$ layer.  
At large moir\'e wavelength, 
moir\'e bands are formed by electron tunneling $t$ between adjacent potential minima, which are well described by a simple tight-binding model on an emergent lattice. The inclusion of the Coulomb interaction between electrons leads to a Hubbard model description. As a hallmark of Hubbard model physics, Mott insulating states are found in  angle-aligned WSe$_2$/WS$_2$ \cite{regan2020mott,tang2019wse2} and twisted AB-homobilayer WSe$_2$ \cite{https://doi.org/10.48550/arxiv.2202.02055} at the filling of $n=1$ hole per moir\'e unit cell. 

One of the fundamental features of the Hubbard model is the local moment formation driven by the on-site repulsion $U$. The presence of local moments in WSe$_2$/WS$_2$ has been observed by measuring the dependence of optical circular dichroism on the magnetic field \cite{tang2019wse2}. It is found that the exciton Zeeman splitting, which is  directly related to the magnetization, saturates above a certain field where the spins are fully polarized. The saturation field depends on the filling factor $n$ and reaches the maximum at $n=1$, as expected from the Hubbard model.  


In this work, we study the charge excitations of the Mott insulator in TMD moir\'e superlattices in the presence of a magnetic field. By exactly solving  the problem of the Mott insulator with one doped hole, we find that as the magnetic field is reduced, the fully polarized state becomes unstable to the formation of a spin polaron -- a bound state of a hole and a spin-flip.  The spin polaron has a kinetic origin due to the correlated hopping of  the hole and the spin-flip on the triangular lattice. Importantly, the binding energy of the spin polaron is on the order of the hole hopping amplitude $t$ and has a strong dependence on the center-of-mass momentum $\bm P$, which we determine exactly. Our work establishes spin polaron, a heavy-mass fermion of charge $-e$ and spin $\frac{3}{2}$, as the fundamental charge carrier in hole-doped Mott insulator over a wide range of magnetic fields, which are experimentally accessible. In contrast, the charge carrier in electron-doped Mott insulator is the doublon with charge $e$ and  spin of $\frac{1}{2}$.

The dichotomy between the charge excitations of opposite signs leads to distinct phases that arise upon doping below and above $n=1$. At $n=1+\delta$ ($\delta>0$),  metallic (Nagaoka) ferromagnetism is favored by the kinetic motion of doublons. At $n=1-\delta$,  a strange metallic state is formed by the dilute Fermi gas of spin polarons with incomplete spin polarization and a gap to adding or removing a charge carrier. As a direct manifestation of the electron-hole asymmetry, we predict a discontinuous jump of the saturation field across $n=1$.  
We further propose compressibility measurements for  detecting spin polarons in TMD moir\'e materials directly. Our work reveals doping-induced itinerant magnetic states in semiconductor moir\'e systems, whose energy scale is defined by the kinetic energy much larger than the exchange interactions. 

After the initial version of this work was completed, we became aware of  the early work~\cite{PhysRevB.97.140507} which identified spin polaron in the context of superconductivity in an extended Hubbard model on the triangular lattice. Related physics in the context of ultracold atoms has been studied using $t-J$ model \cite{https://doi.org/10.48550/arxiv.2106.09600}.   Compared to these studies, our work not only introduces TMD moire superlattices as the promising material platform for the realization of spin polaron, but also identifies its experimental manifestation, namely the dependence of the charge gap on the magnetic field. 

{\it Hubbard model description and the Mott insulator at $n=1$.---}
The starting point for our analysis of a TMD moir\'e heterobilayer under a magnetic field is the canonical Hubbard model on  a triangular lattice  \cite{PhysRevLett.121.026402}: 
\begin{eqnarray} \label{H_Hubbard}
H = - t \sum_{\langle i,j \rangle} (c^\dagger_i c_j + h.c. ) + U \sum_i n_{i \uparrow}n_{i \downarrow} 
+  \frac{h}{2} \sum_i \lp n_{i \uparrow} - n_{i \downarrow}   \rp \nonumber.\\ 
\end{eqnarray}
$c^\dagger_i$ is the creation operator of a doped charge in the moir\'e superlattice. For simplicity of presentation, we assume the doped charge is of electron type.  As we discuss later, the long-range Coulomb interaction does not affect the formation of spin polaron in the limit of large Hubbard $U$.
For typical TMD moir\'e materials, $t\sim 1$~meV \cite{PhysRevB.102.201115} is much smaller than the on-site Coulomb repulsion $U$, leading to the strong-coupling regime of the Hubbard model. 


At half-filling ($n=1$), the Mott insulator is a quantum antiferromagnet governed by the spin-$\frac{1}{2}$  Heisenberg model on the triangular lattice: $H_J=J\sum_{\langle i j\rangle} {\bm s}_i \cdot {\bm s}_j$, with $J=4t^2/U$ and 
${\bm s}$ is the spin-$\frac{1}{2}$ operator.   %
Since $t\ll U$, the antiferromagnetic exchange interaction $J$ in TMD moir\'e superlattices is generally weak \cite{PhysRevB.104.214403}. As a result, the antiferromagnetic Mott insulator becomes fully polarized above a small saturation field $h_s^0$ whose value  is set by $J$ (see  SM):
$
h_s^0 = \frac{9}{2} J = 18 \frac{t^2}{U},
$
where the superscript `0' refers to an undoped Mott insulator. For example, using $t=1$ meV and assuming $U\sim 50$~meV for angle-aligned WSe$_2$/WS$_2$, $J$ is only $0.08$~meV and the corresponding saturation field is $1$~T (using the $g$-factor $6.7$ for holes in WSe$_2$). This value is comparable to the saturation field measured by MCD at $n=1$ \cite{tang2019wse2}. 

Thanks to the narrow bandwidth, the full magnetization curve of TMD moir\'e materials can be measured over the entire range of filling factors $0 \leq n \leq 2$, which is not possible elsewhere. The ability to achieve full spin polarization in doped Mott insulators opens access to rich and previously unexplored Hubbard model physics on the triangular lattice, as we shall show below.      

\begin{figure}[t] 
	\includegraphics[width= 0.85\columnwidth]{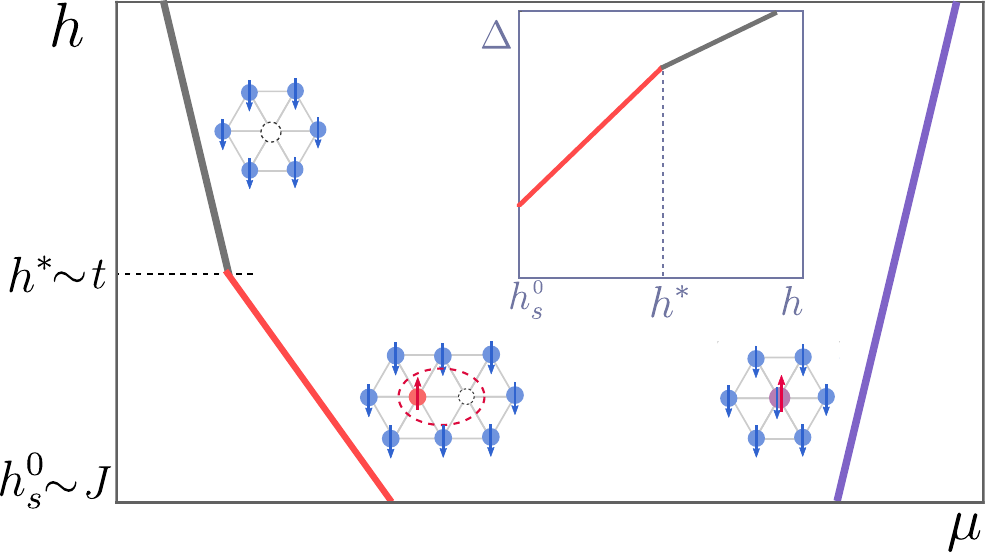}
	\caption{ The gap edges of the Mott insulator as a function of the magnetic field $h$. The upper and lower edges of the gap are defined by the energy cost of adding a charge $e$ and $-e$ quasiparticle, respectively. The charge $e$ quasiparticle is a $s=\frac{1}{2}$ doublon. The charge $-e$ quasiparticle transitions from a $s=\frac{1}{2}$ hole to a $s=\frac{3}{2}$ spin polaron at $h=h^*= \varepsilon_b - E_{sw}^0 \sim t$, resulting in a change of slope in the lower gap edge and the charge gap at $n=1$ (inset).} 
	 \label{fig:excitation_energy}
\end{figure}

{\it Charge exitations in Mott insulator.---}
As a first step towards the study of doped Mott insulators, we consider charge excitations of the Mott insulator at $n=1$ at full spin polarization induced by a magnetic field $h>h_s^0$.   
 Interestingly, we find that the charge $e$ and $-e$ excitations have very different nature, as represented in Fig.~\ref{fig:excitation_energy}.        

Charge $e$ excitation is simply a doublon created by adding an electron with minority spin, which costs a minimum energy 
\begin{eqnarray}
E_d = E_d^0 + \frac h 2 = U - \mu - 6 t  + \frac{h}{2},
\end{eqnarray} 
where $E_d^0$ is the minimum energy of the doublon in the absence of magnetic field, $\mu$ is the chemical potential, $-6t$ comes from the kinetic energy of the added electron  at the bottom of the band at ${\bm k}=0$, and $h/2$ comes from the Zeeman energy of the added minority spin. 
The nature of charge $-e$ excitation depends on the magnetic field $h$. When $h$ is sufficiently large, the lowest energy excitation is simply a hole with a minimum energy given by
\begin{eqnarray}
E_h = E_h^0 + \frac h 2 = \mu - 3t  + \frac{h}{2},
\end{eqnarray}
where $E_h^0= \mu -3t$ is the minimum energy of the hole in the absence of magnetic field, which comes from the kinetic energy of the hole at the band maxima $\bf k= \pm K$.  

However, when the magnetic field is reduced below a certain value $h^*$ (with  $h^*\gg h_s^0$ for $t \gg J$; we assume well-separated scales of energies here for clarity, and address more realistic parameters later on), 
we find that the lowest energy state of the Mott insulator with one hole is no longer fully spin polarized, but contains one spin flip that is bound to a hole.   
The bound state of the hole and the spin-flip is a spin polaron, a composite quasiparticle carrying spin $s =\frac{3}{2}$ along the field direction.  $h^*$ is the saturation field for the Mott insulator with one hole. Viewed from a complementary perspective, $h^*$ is the dividing line between the domains with two types of charge $-e$ excitations in the Mott insulator: the bare hole and the spin polaron.

As we show below, the origin of the spin polaron formation is purely kinetic. The spin flip gains kinetic energy of the order of $t \gg J $ by exchanging its position with an adjacent hole.
Remarkably, this highly restricted kinetic process is sufficient to bind them together on the triangular lattice,  but not on square or honeycomb lattices. 
The result is an {\it itinerant} spin polaron  whose binding energy depends on its center-of-mass momentum $\bm P$. We find that at $h<h^*$, the  energy cost of adding a carrier with charge $-e$ is

\begin{eqnarray}
\begin{split}
E_{sp} &= E_h  + E_{sw}  - \varepsilon_b \\
& = E_h^0 + E_{sw}^0 + \frac{3h}{2} - \varepsilon_b,
\end{split}
\end{eqnarray}
where $E_{sw}^0$ is the minimum kinetic energy of spin-waves in the absence of magnetic field and $\varepsilon_b \sim t $ is the binding energy of the spin polaron at zero momentum $\bm P=0$. The total Zeeman energy $\frac{3h}{2}$ comes from  the $s =\frac{3}{2}$ of the spin polaron. 

Comparing the expressions for $E_{sp}$ and $E_h$, we see that the spin polaron has lower energy than a hole at $h<h^*$ with $h^* = \varepsilon_b - E_{sw}^0$.  At large Hubbard $U$, the binding energy $\varepsilon_b \sim t $ is significantly larger than $E_{sw}^0 \sim J = 4 \frac{4t^2}{U}$.      In a  wide range of fields $h_s^0 \sim J < h < h^* \sim t$, spin polarons are the lowest-energy charge carriers upon hole doping of the Mott insulator.  Note that our spin polaron exists on top of  the field-polarized state of the Mott insulator, which is fundamentally different from the magnetic polaron in quantum antiferromagnets at $h=0$    \cite{PhysRevB.39.6880,RevModPhys.78.17,PhysRevB.39.12232,PhysRevLett.60.2793,PhysRevB.38.8879,PhysRevLett.60.944}. For example, since the noncollinear antiferromagnetic state on the triangular lattice spontaneously breaks spin rotational symmetry, the magnetic polaron at $h=0$ does not have a well-defined spin quantum number, in contrast with the $s=\frac{3}{2}$ spin polaron we find here.   

The field-induced transition in the type of the charged excitations is reflected in the charge gap of the Mott insulator, defined as  $\Delta = E_{+e} + E_{-e}$: 
\be
\Delta(h) = \left\{\begin{matrix}
\Delta_0 + h, \quad \quad \quad h>h^*,
\\ 
\Delta_0 + E_{sw}^0 - \varepsilon_b + 2h, \ \ h_s^0< h <h^*,
\end{matrix}\right.  
\ee
where $\Delta_0 $ is  field-independent. 
Due to the different spin quantum numbers of the hole and the spin polaron, $\Delta(h)$ shows a change of slope at $h^*$, as illustrated in Fig.~\ref{fig:excitation_energy}.

Remarkably, because of its purely kinetic origin, the spin polaron appears already in the limit $U= \infty$ (\cite{PhysRevB.97.140507}, see also \cite{https://doi.org/10.48550/arxiv.2209.05398,https://doi.org/10.48550/arxiv.2106.09600}).
In what follows, we start by considering $U = \infty$ first ($t/U = 0$). Next, we study the spin polaron formation and the saturation field at finite $t/U$, showing that surprisingly, the binding between the hole and the spin flip is further \emph{enhanced} at finite $t/U$.  We finally conclude by discussing experimental signatures of the spin polaron.




\begin{figure}[t] 
	\includegraphics[width= 1\columnwidth]{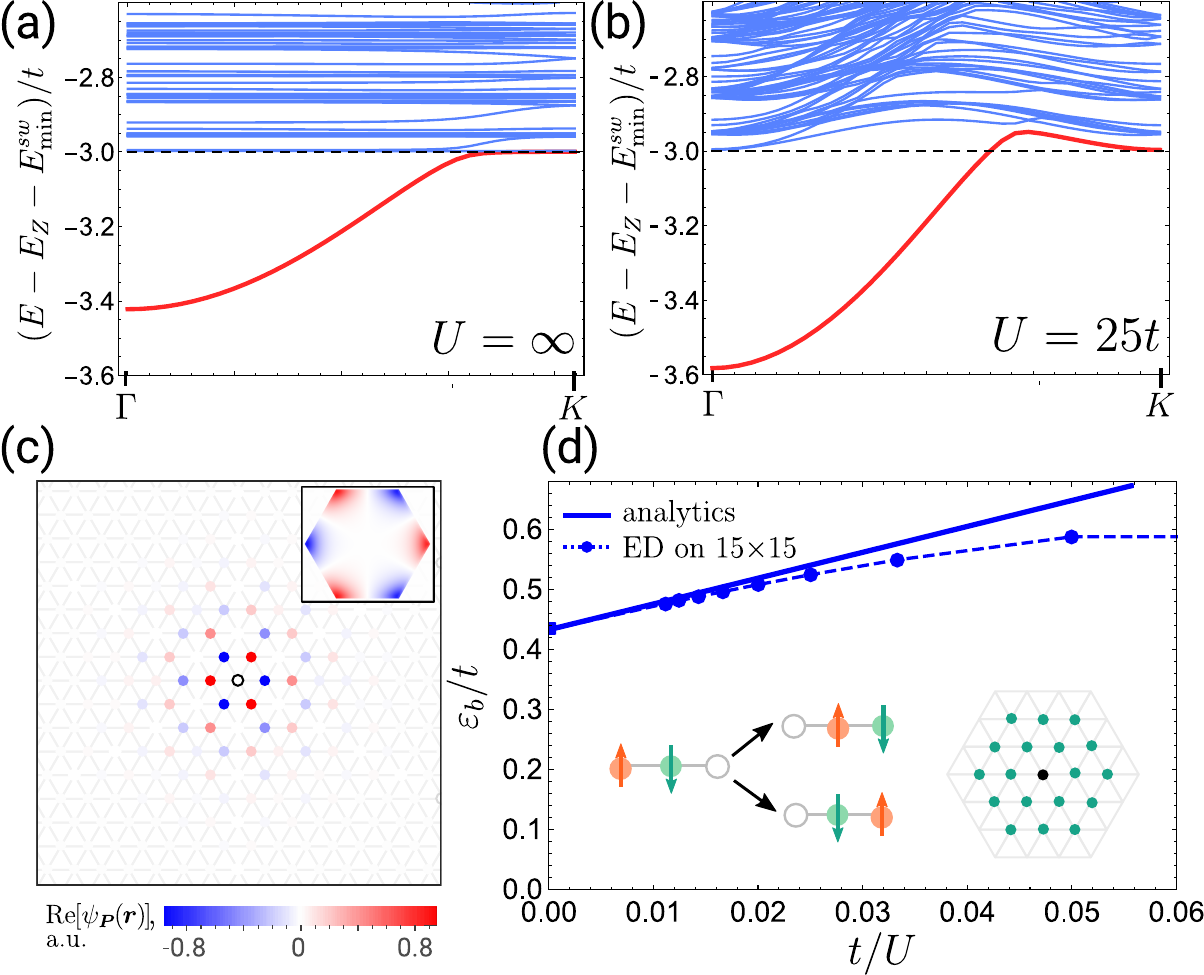}
	\caption{  The energy spectrum of the Hubbard model at $n=1$ doped with one hole as a function of the center of mass momentum $\bm P$ along $\Gamma-K$ direction at  (a) infinite $U$ and (b) $U=25t$.  The dispersive bound state (red line) is found below the continuum spectrum (blue); the total Zeeman energy of the state and the minimum  energy of spin waves are $E_Z = \frac{3}{2}h$ and $E_{sw}^{\text{min}} =18 \frac{t^2}{U} = \frac{9}{2} J$. (c) The real space  wavefunction in relative coordinates for  $\bm P = \bm 0$ at $U = \infty$. Inset: momentum-space wavefunction in relative coordinates.  (d) The the binding energy of the spin polaron increases as a function of $t/U$. The results of the exact diagonalization of the full Hubbard model on $15\times15$ lattice and of the analytical approach at order $t^2/U$ are shown by dots and solid line, correspondingly. 
	} \label{fig:numerics1_main}
\end{figure}

{\it Spin polaron.---}
Let us first study the spin polaron in the limit $U \rightarrow \infty$, where the physical picture becomes especially simple. A single hole or doublon with momentum $\bm k$ will have energy $\varepsilon_{h/d} (\bm k) = \frac{h}{2} \pm t \gamma_{\bm k}$ with  $\gamma_{\bm k} = \sum_n \lp e^{i \bm k \cdot \bm t_n} + e^{-i \bm k \cdot \bm t_n}\rp$, where $\bm t_{n=1,2,3}$ are the three basis vectors on  triangular lattice. We examine the state containing one hole and one spin flip using the general ansatz: 
\be \label{ansatz}
\ket{\psi_h} = \sum_{n,m} \alpha_{nm} c_{n \downarrow} S^+_m \ket{FM_{n=1}}.
\ee
 Here, $S_m^+ \equiv c^\dagger_{m \uparrow} c_{m \downarrow}$.  The vacuum state corresponds to a fully-polarized state with single occupancy at each site
$\ket{FM_{n=1}} = \prod_i c_{i\downarrow}^\dagger \ket{0} 
$. The wavefunction must necessarily vanish at the origin, reflecting the fact that the positions of a spin flip and a hole cannot coincide.

%
The Hubbard Hamiltonian at $U=\infty$, which forbids double occupancy, acting on eq.~\eqref{ansatz} reduces to a two-particle problem. This problem can be separated into center of mass and relative motion, which in relative coordinates  becomes a version of a tight-binding model on  a triangular lattice. 
The details of the calculation are provided  in Supplemental Material \cite{SM}. 
We find that at $\bm P = 0$, the bound state of a hole and a  spin flip occurs for one of the inversion-odd representations of the group $D_6$ and follows from the especially simple self-consistency equation 
\be
1 + \sum_{\bm q} \frac{2 t \sin \bm q  \cdot \bm t_1 \lp \sin \bm q \cdot \bm t_1 + \sin \bm q \cdot \bm t_2 + \sin \bm q \cdot \bm t_3\rp }{ E - \frac{3}{2}h -  t\gamma_{\bm q}} = 0.
\ee
This produces a bound state (spin polaron) with energy $E_{sp} (\bm P =0) = E_h -\varepsilon_b^{(0)}  + h $, where the binding energy is found to be  $\varepsilon_b^{(0)} \equiv  \varepsilon_b (\bm P=0)  \approx 0.42 t$ and the $+h$ contribution is the energetic cost of a spin flip. The fact that the binding energy is proportional to $t$ 
indicates the  kinetic origin of spin polaron formation as we discussed above. 

Next, we solve the tight-binding equation describing the relative motion of the hole and the spin flip  with a  finite center of mass momentum in order to find the spin polaron dispersion. The spectrum for $\bm P$ along the $\Gamma-K$ direction obtained from  exact diagonalization of the tight-binding equation on a lattice of 866 sites with periodic boundary conditions is shown in Fig.~\ref{fig:numerics1_main}(a). The dispersive bound state is found below the band bottom.  We find that the mass of the bound state is $m_{sp} \approx 13 m_h$, where the mass of the bare hole is $m_h = \frac{2}{3}\frac{1}{t a^2}$. 
Fig.~\ref{fig:numerics1_main}(c) shows the real-space wavefunction of the spin polaron in the relative coordinates at $\bm P = 0$ . The spin polaron is tightly bound on a lengthscale of the order of one lattice spacing and the wavefunction realizes the one-dimensional antisymmetric irrep $\Gamma_3$ of the dihedral group $D_6$ and vanishes exactly at the origin.  As seen both in Fig.~\ref{fig:numerics1_main}(a), the state merges with continuum at $\bm P = K$. 

For a single doublon, we find, both analytically using the approach described above  and numerically (see SM), that the spin polaron does not form. It similarly does not form on the square lattice, for either doping. In particular, this is dictated by the symmetry of the solution: the wavefunction of the bound states must vanish at the origin in the relative coordinates, i.e. $\alpha_{nn} = \beta_{nn}= 0$. The single-particle spectrum of a doped electron on the triangular lattice, or doped electron/hole on a  square lattice has only one band minimum and therefore, the low-energy states of such excitations cannot have a node. In contrast, on the triangular lattice, the hole dispersion has two band minima at $\pm K$  points. An antisymmetric superposition of $\pm K$ states, shown in the inset in Fig.~\ref{fig:numerics1_main}(b), allowing for the existence of a spin polaron.  

In the case of charge-transfer insulator described by multi-band Hubbard models \cite{PhysRevB.102.201115,zhang2021electronic}, the dispersion of the charge carriers doped below the fully polarized state at $n=1$ still has two band minima at $\pm K$, which leads to the spin polaron formation. The situation will be different for electron doping, which we leave to a future study.

\textit{Finite U. ---}
We now consider the effect of large finite $U$ (small nonzero $t/U$). 
The effective Hamiltonian at the order $t^2/U$ includes not only the spin exchange, but also the correlated hopping. The correlated hopping comes from the second-order processes wherein the spin or the hole can move over one or two sites (see the inset in Fig.~\ref{fig:numerics1_main}(d)). Importantly, these processes only occur when the hole and the spin flip are in the vicinity of each other. While the correlated hopping is commonly ignored in the literature\cite{https://doi.org/10.48550/arxiv.2209.05398,https://doi.org/10.48550/arxiv.2106.09600} , we show that these microscopic kinetic processes can have important effects on  spin polaron formation.

We obtain the full analytical solution for the bound state problem at the order $\frac{t^2}{U}$ (see SM). The spectrum of the spin polaron is shown in Fig.~\ref{fig:numerics1_main}b) at  $U = 25t$. In Fig.~\ref{fig:numerics1_main}(d), we plot  the dependence of the binding energy on $t/U$ as obtained from the analytical approach at the order $\frac{t^2}{U}$. Also shown is  the result of the exact diagonalization of the full Hubbard model Hamiltonian on a $15 \times 15$ lattice in the appropriate spin and charge sectors. The two methods show excellent agreement up to $t/U \approx 0.05$, which corresponds to $J/t \approx 0.2$. 
Remarkably, the binding energy of the spin polaron \emph{increases} with  $t/U$. At large but finite $U$,  the spin flip can become delocalized to lower its kinetic energy, which competes with the formation of the bound state. Nevertheless, the proximity to the hole enables a large number of correlated hopping processes on the triangular lattice, which leads to an additional gain in kinetic energy.  This increases the binding energy of the spin polaron and dominates over the spin delocalization. Thus, correlated hopping enhances the stability of spin polarons at finite $U$, an effect which has been overlooked before\cite{PhysRevB.97.140507}. In contrast, neglecting correlated hopping, i.e. working with the $t-J$ model, will produce a decrease in the binding energy with $t/U$ (see Fig.~S8 in the SM), which is incorrect.


{\it Finite doping.---} We now consider the case of finite doping density. The  Hubbard model on triangular lattice has been extensively studied in the absence of magnetic field. It is known that, at  $U \gg t$ and for the electron doping ($n>1$), the Nagaoka ferromagnetic state  \cite{PhysRev.147.392,https://doi.org/10.1002/andp.19955070405,PhysRevB.41.2375} arises due to the kinetic energy gain of the doublons, which dominates over the weaker antiferromagnetic exchange interaction between localized spins $J=4t^2/U \ll t$. In contrast, for any amount of doped holes, the ferromagnetic state is unstable at zero magnetic field  \cite{PhysRevB.40.9192,richmond1969ferromagnetism,PhysRevB.41.2375,PhysRevLett.95.087202,https://doi.org/10.1002/andp.19955070405}, while the nature of the true ground state is hard to determine.  

Our results on charge excitations of the Mott insulator under a magnetic field provide new insight. As we have shown, while the undoped Mott insulator is already fully polarized at small magnetic fields above $h^0_s \sim J \propto t^2/U$, 
the state with one hole can only achieve full polarization above a  larger field $h^*\sim t > h_s^0$, at which the first spin flip appears that is bound to the hole. Now consider a finite but small density $\delta$ of holes. At high field, the fully polarized state is a dilute Fermi gas of holes. As the field is reduced, provided that the hole density is  sufficiently low,  the first spin flip to appear should also bind with one hole. 
It follows from this argument that the saturation field $h_s$ at finite hole density should approach $h^*$ as $\delta \rightarrow 0$. 
In contrast,  upon electron doping, Nagaoka mechanism eventually leads to ferromagnetic ground state at zero external field (or immediately becomes ferromagnetic in the limit $U = \infty$). Note that $h^*$ remains finite even when $U=\infty$, whereas $h^0_s=0$ in this limit. Therefore, we conclude that  the saturation field as a function of doping shows a  {\it discontinuous} jump from $h_s= h^*$ at $n=1^-$ to $h_s^0 < h^*$ at $n=1^+$.

\begin{figure}[t] 
	\includegraphics[width= 1.0\columnwidth]{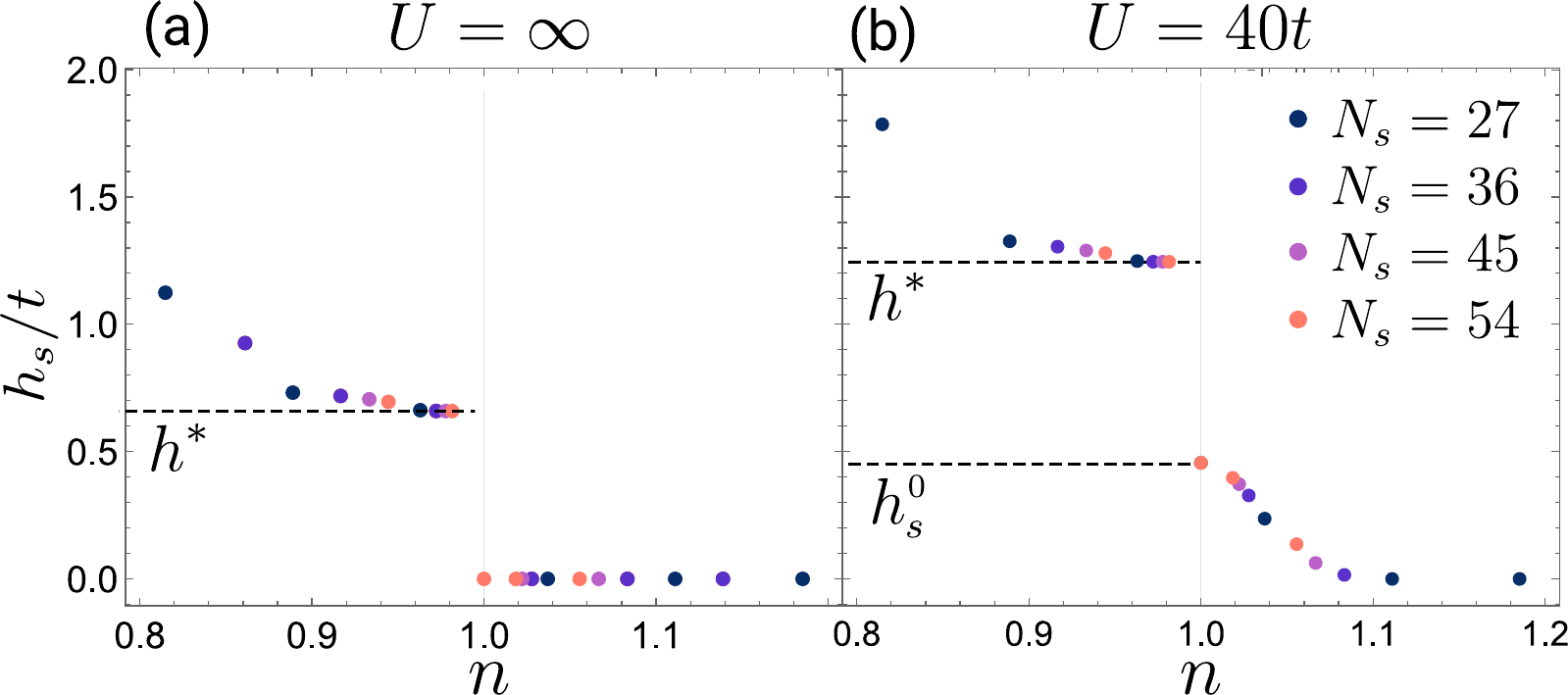}
	\caption{ The saturation field $h_s$ at different fillings obtained by exact diagonalization of the Hubbard  model with an odd number of doped electrons/holes on $3\times L$ geometry with  $L=9, 12, 15, 18$ in the sector with a single spin flip. Panel (a) represents  the limit $U = \infty$ and panel (b) corresponds to $U = 40 t$.  The dashed black line shows the field $h^*$ for the undoped Mott insulator at $n=1$.
	} \label{fig:ED}
\end{figure}

This conclusion is supported by our calculation of the saturation field $h_s$ as a function of doping, using exact diagonalization of the Hubbard model shown Fig.~\ref{fig:ED}. The calculations were performed at fixed number of holes $N_h = 1,3,5$ on $3$-leg ladders with periodic boundary conditions and various lengths $L=9, 12, 15, 18$. By comparing the energy of the state with a single spin flip to that of a fully polarized state, we obtain a lower bound on the saturation field. 
Fig.~\ref{fig:ED} shows the results of exact diagonalization  for infinite-$U$ Hubbard model and at $U = 40 t$.  On the hole doping side, the field $h_s$ approaches $h^*$ (its value is enhanced because of the finite-width effects, and fully agrees with our tight-binding calculation). The parameters that we chose are realistic for many TMD  moir\'e materials\cite{PhysRevB.102.201115}, where a large window of magnetic fields exists for the spin polaron predicted here to be observed. On the electron doping side the saturation field equals $h_s^0$ at $n \rightarrow 1^+$ and exhibits behaviour expected from a Nagaoka ferromagnet at infinite and finite Hubbard $U$, as seen in panels (a) and (b).

Finally, we discuss the effect of the long-range Coulomb repulsion $\sum V_{ij} n_i n_j$ on the binding energy of  spin polarons. At $U = \infty$, it does not affect the energy of a single spin polaron, because the system contains only one hole and its Coulomb energy is independent of the spin configuration.  At finite but large $U$ there will be a small correction of the order of $\frac{t^2}{U} \frac{V}{U} \ll t$ due to the small amplitude of admixing doublons. 

{\it  Experimental implications.---}
The particle-hole asymmetry of the saturation field, especially its discontinuity at $n=1$, reflects the distinction between the doublon and the spin polaron in doped Mott insulator on the triangular lattice. In light of our theory, it is encouraging to note that the saturation field in WSe$_2$/WS$_2$ measured at $T=1.7$~K \cite{tang2019wse2} indeed decreases with doping at $n>1$, increases with doping at $n<1$, and shows a large rapid change across $n=1$, which we expect will sharpen into a discontinuity at $T=0$.

The presence of spin polaron  can be established by the dependence of the lower edge of the Mott gap at $n=1$ on the magnetic field, which can be obtained from compressibility measurements. As shown in Fig.~\ref{fig:excitation_energy}, our theory predicts a linear dependence of the lower gap edge on the field with a change in the slope by a factor of 3 at $h=h^*$, which shows the different spin quantum numbers: $s=\frac{3}{2}$ for the spin polaron and  $s=\frac{1}{2}$ for the bare charge carrier below $n=1$.   

Our theory predicts that at small hole doping, a Fermi liquid of $s=\frac{3}{2}$ spin polarons can form in a range of magnetic fields below $h^*$ and above $h_s^0$. This is a pseudogap metallic state with heavy fermion mass  that has a gap to adding an $s=\frac{1}{2}$ electron/hole, and also exhibits filling-dependent magnetization plateaux. Its detailed study appeared in ref.~\cite{https://doi.org/10.48550/arxiv.2209.05430}.

\textit{Note added. ---} Recent measurements of electronic compressibility  in twisted double bilayer WSe$_2$~\cite{https://doi.org/10.48550/arxiv.2206.10631}  revealed a kink in the charge gap  as a function of magnetic field, consistent with our theory of the transition between the spin polaron and the bare hole quasiparticles.

This work was supported by the Air Force Office of Scientific Research (AFOSR) under award FA9550-22-1-0432 and the David and Lucile Packard Foundation.

\clearpage

\setcounter{equation}{0}
\setcounter{figure}{0}
\setcounter{table}{0}
\makeatletter
\renewcommand{\theequation}{S\arabic{equation}}
\renewcommand{\thefigure}{S\arabic{figure}}

\newpage
\clearpage
\pagebreak
\onecolumngrid
\begin{center}
\textbf{\large Supplemental Materials}
\end{center}

Our starting point is the Hubbard Hamiltonian in two dimensions in magnetic field:
\be \label{H_SM}
H = U \sum_i n_{i \uparrow}n_{i \downarrow} +  \frac{h}{2} \sum_i \lp n_{i \uparrow} - n_{i \downarrow}  \rp - t \sum_{\langle i,j \rangle} \lp c^\dagger_i c_j + h.c. \rp
\ee
Even though we are interested in the case of triangular lattice, we do not specify the type of the lattice until later in order to make comparison with the case of square lattice.

\section{Half filling}

At half filling, the effective Hamiltonian is $H_{eff} = \frac{ t^2}{U}\sum_{\langle i,j \rangle} \lp \bm S_i \cdot \bm S_j - 1\rp + \frac{h}{2}  \sum_i S_{zi} $, which corrseponds to the 2D Heisenberg antiferromagnet in magnetic field (note that we defined $S_{zi} = n_{i \uparrow} - n_{i \downarrow}$).  Consider large field limit first, $h> h_c$, where $h_c$ is the saturation field. Then the excitations are gapped and we can assume that the spin-wave density is low. We can expand the Holstein-Primakoff transformation to all orders and collect the terms that only live in $0,1$-boson sectors:
\be
\begin{split}
S_i^+ &= a^\dagger_i \lp 1 - a_i^\dagger a_i \rp \\
S_i^- &=  \lp 1 - a_i^\dagger a_i \rp a_i
\end{split}
\ee
The commutation relation gives
\be
\begin{split}
  [S^+_i,S^-_i] = S_{zi} -3 (a^\dagger_i)^2 (a_i)^2
\end{split}
\ee
where $S_{zi} = n_{i \uparrow} - n_{i \downarrow} = 2 a_i^\dagger a_i -1$.

The Hamiltonian is expressed through bosonic degrees of freedom:
\be
H_{eff}|_{\nu = \frac{1}{2}} = \frac{t^2}{U} \sum_{\langle i,j\rangle} \lp 2 (a_i^\dagger a_j  - a_i^\dagger a_i ) - 2 (a_i^\dagger a_i^\dagger a_i a_j + a_i^\dagger a_j^\dagger a_j a_j - a_i^\dagger a_j ^\dagger a_i a_j )  \rp + \langle i\Leftrightarrow j\rangle + \frac{h}{2} \sum_i (2 a_i^\dagger a_i -1)
\ee
This is the Hamiltonian that we can use to study  the tranisiton from the `ferromagnetic' to the ferrimagnetic state the occurs when the gap closes and the bosons condense. The single-particle spectrum is determined from 
\be
\begin{split}
H_{1pt}|_{\nu = \frac{1}{2}} &= \frac{2 t^2}{U} \sum_{\langle i,j\rangle} \lp a_i^\dagger a_j  - a_i^\dagger a_i  \rp + \langle i\Leftrightarrow j\rangle + \frac{h}{2} \sum_i (2a_i^\dagger a_i -1) = \\ 
&= \sum_{\bm k } \lp \frac{2 t^2}{U}  \sum_{n=1}^z (e^{i \bm k \cdot \bm t_{n}}-1)  + h \rp a^\dagger _{\bm k} a_{\bm k} + const,
\end{split}
\ee
where $\bm t_n$ are the vectors pointing from a site to the nearest neighbors. Plugging that in, we obtain that the gap closes at $h_c = 2 z \frac{t^2}{U}$ for bipartite lattices (at $\bm k_0 = (\pi,\pi)$ for square lattice) and $h_c = 18 \frac{t^2}{U}$ for triangular lattice ( at $K, K'$ points).
Below the critical value of the magnetic field $h_c$, the interacting bosons condense.

\section{Bound state of a spin flip and a hole in a ferromagnetic background}

First, we project the Hamiltonian onto the subspace forbidding double occupancy. To the leading order in $t$, the effective Hamiltonian becomes:
\be
H_{eff} =   \frac{h}{2} \sum_i \lp n_{i \uparrow} - n_{i \downarrow}\rp -t\sum_{\langle i,j \rangle, \sigma}  \lp  c^\dagger_{j \sigma} c_{i \sigma}  P_{j,\overline{\sigma}} P_{i,\overline{\sigma}}   + h.c. \rp 
\ee
For the purposes of this section, it is easier if we rewrite it more explicitly:
\be
\begin{split}
H_{eff} &= H_h + H_{t}^{\uparrow} + H_t^{\downarrow} =  \\
&=\frac{h}{2} \sum_i \lp n_{i \uparrow} - n_{i \downarrow}\rp 
-t\sum_{\langle i,j \rangle}  \lp c_{j\uparrow}^\dagger c_{i\uparrow}(1-n_{i\downarrow})(1-n_{j\downarrow})  + h.c. \rp -t\sum_{\langle i,j \rangle}  \lp c_{j\downarrow}^\dagger c_{i\downarrow}(1-n_{i\uparrow})(1-n_{j\uparrow})  + h.c. \rp 
\end{split}
\ee
Let us introduce the vacuum state:
\be
\ket{vac} = \prod_i c_{i\downarrow}^\dagger \ket{0}
\ee
Let us find the eigenstate of the Hamiltonian which has exactly one hole and one spin-flip excitation in a half-filled ferromagnetic background. We use the ansatz:
\be \label{ansatz_SM}
\ket{\psi} = \sum_{n,m} \alpha_{nm} c_{n \downarrow} S^+_m \ket{vac}
\ee
Thus, $\alpha_{nm}$ is the amplitude of having a hole at position $n$ and a spin-up at position $m$ in state $\ket{\psi} $. We must set $\alpha_{nn} = 0$. 

It is obvious that
\be
H_h \ket{\psi} = \frac{3}{2}h \ket{\psi}
\ee
After a straightforward but lengthy calculation, we find:
\be
H^\uparrow_t \ket{\psi} = t \sum_{\langle i,j\rangle}   \alpha_{ij} c_{j\downarrow}S^+_i  +\alpha_{ji} c_{i\downarrow}S^+_j  \ket{vac}
\ee
which has a clear physical meaning: this is the hopping that occurs only when hole and spin-flip are neighboring.

Similarly,
\be
H^\downarrow_t \ket{\psi} = t \sum_{\langle i,j \rangle} \lp c_{j\downarrow} \sum_m \alpha_{im} S^+_m \ket{vac} + c_{i\downarrow} \sum_m \alpha_{jm} S^+_m \ket{vac}\rp 
\ee
which tells us that hole hops onto a spin-down place only if that place is not occupied by a spin-flip, and the spin flip can be anywhere but on the respective neighboring site. 

Collecting everything, and projecting onto one of the orthogonal states in the suprtposition, we find that the relation for  the eigenstate is
\be
\frac{3}{2} h \alpha_{\ell s} + t \sum_n \alpha _{s \ell } (\delta(\ell-s - \bm t_n) + \delta(\ell-s + \bm t_n))+ t \sum_n (\alpha _{\ell + \bm t_n, s} + \alpha _{\ell- \bm t_n, s}) = E \alpha_{\ell s}
\ee
Next, we separate the coordinates into the one of center of mass and the relative one: $\alpha_{nm} = \alpha(\bm R + \bm r, \bm R)$, where $\bm R = \bm r_m$ and $\bm r = \bm r_n - \bm r_m$.

This allows us to rewrite the last equation as
\be \label{eq:alpha}
\frac{3}{2} h \alpha(\bm R + \bm r, \bm R) + t \alpha(\bm R , \bm R + \bm r) \sum_n (\delta(\bm r - \bm t_n) + \delta(\bm r + \bm t_n) ) + t \sum_n (\alpha(\bm R + \bm r + \bm t_n, \bm R) + \alpha(\bm R + \bm r - \bm t_n, \bm R)) = E \alpha(\bm R + \bm r, \bm R)
\ee

We use $\alpha( \bm R + \bm r, \bm R)  =  \sum_{\bm P}  \psi_{\bm P}(\bm r)  e^{i \bm P \cdot \bm R}$ in order to find solution with specific center of mass momentum:
\be
\frac{3}{2} h \psi_{\bm P} (\bm r) + t \sum_n \psi_{\bm P} (-\bm r)  \lp e^{i \bm P \cdot \bm t_n}  \delta (\bm r - \bm t_n) + e^{-i \bm P \cdot  \bm t_n } \delta (\bm r + \bm t_n) \rp  + t \sum_n \lp  \psi_{\bm P} (\bm r + \bm t_n) + \psi_{\bm P} (\bm r - \bm t_n)\rp = E \psi_{\bm P} (\bm r)
\ee
Which is more conveniently grouped as 
\be \label{eq:psi}
\lp E - \frac{3}{2} h \rp  \psi_{\bm P} (\bm r) - t \sum_{\pm t_n}  \psi_{\bm P} (\bm r + \bm t_n)  = t \sum_{\pm t_n}  \psi_{\bm P} (-\bm r) e^{i \bm P \cdot \bm t_n} \delta (\bm r - \bm t_n)
\ee
And is solved together with the condition $\psi_{\bm P}(0) = 0$.
Let us perform another Fourier transform $ \psi_{\bm P} (\bm r) =  \sum_{\bm k}  \varphi_{\bm P} (\bm k)  e^{i \bm k \cdot \bm r}$, which leads to 
\be  \label{master_eq}
\lp E - \frac{3}{2} h  - t \gamma_{\bm k}   \rp \varphi_{\bm P} (\bm k)= \lambda_{\bm P} (\bm k)
\ee
where $\gamma_{\bm k} = \sum_n e^{i \bm k \cdot \bm t_n} + e^{-i \bm k \cdot \bm t_n}$ (where $\bm t_n$ are the two \textit{basis} vectors of the square lattice or the three for triangular one), and 
\be
\lambda_{\bm P}(\bm k)   = t \sum_{\pm \bm t_n}  \psi_{\bm P} (\bm t_n) e^{i ( \bm k - \bm P) \cdot \bm t_n}  
\ee
From here, there are two options:

(a) Either the condition 
\be  \label{eq:zero}
\lambda_{\bm P}(\bm k) = t \sum_{\pm \bm t_n}  \psi_{\bm P} (\bm t_n) e^{i ( \bm k - \bm P) \cdot \bm t_n}      = 0
\ee
is true. In this case, 
\be
 E(\bm k) = \frac{3}{2} h  + t \gamma_{\bm k}  
\ee

And $\varphi_{\bm P} (\bm k)$ are only restrained by the condition \eqref{eq:zero} and $\psi_{\bm P }(0) = 0$. This corresponds to an independent motion of hole and the spin wave, which are in this case not bound; 

or  (b):
\be  \label{eq:nonzero}
\lambda_{\bm P}(\bm k) = t \sum_{\pm \bm t_n}  \psi_{\bm P} (\bm t_n) e^{i ( \bm k - \bm P) \cdot \bm t_n}    \neq 0
\ee
In this case, the eigenstates states can be found as 
\be \label{master_sol}
 \varphi_{\bm P} (\bm k)= \frac{ \lambda_{\bm P} (\bm k)  }{ E(\bm P) - \frac{3}{2} h  - t \gamma_{\bm k}   }
\ee

and the energy $E = E(\bm P)$ and is found from the self-consistent condition. Let us derive the condition by plugging  $\psi_{\bm P} (- \bm t_n) =  \sum_{\bm q}  \varphi_{\bm P} (\bm q)  e^{- i \bm q \cdot \bm t_n}$ in the following expression
\be
\sum_{\pm \bm t_m} \psi_{\bm P } (- \bm t_m) e^{i (\bm P - \bm k) \cdot \bm t}=\sum_{\pm \bm t_m} \sum _{\bm q} \varphi_{\bm P}(\bm q) e^{i (\bm P - \bm k - \bm q) \cdot \bm t_m}  =  \sum_{\pm \bm t_m} \sum _{\bm q}  e^{i (\bm P - \bm k - \bm q) \cdot \bm t_m}  \frac{t \sum_{\pm \bm t_n} \psi_{\bm P} (- \bm t_n) e^{i(\bm P - \bm q)\cdot \bm t_n}}{E(\bm P) - \frac{3}{2}h - t \gamma_{\bm q}}
\ee
where in the second equality, we used eq.~\eqref{master_sol}. Let us define $\widetilde \epsilon_{\bm P} = \frac{E(\bm P) - \frac{3}{2}h}{t}$, which brings us to:
\be \label{eq:lambda}
\lambda_{\bm P} (\bm k ) =  \sum _{\bm q} \frac{ \gamma_{\bm k + \bm q - \bm P }  \lambda_{\bm P}(\bm q)}{\widetilde \epsilon_{\bm P} -  \gamma_{\bm q}}
 \ee

This integral equation can be rewritten as 
\be
\sum _{\bm q} \lambda_{\bm P}(\bm q) \lp \delta (\bm q - \bm k) - \frac{ \gamma_{\bm k + \bm q - \bm P }  }{\widetilde \epsilon_{\bm P} -  \gamma_{\bm q}} \rp = 0
\ee
which only has nonzero solutions if the self-consistency condition 
\be
\det \lp \delta (\bm q - \bm k) - \frac{ \gamma_{\bm k + \bm q - \bm P }  }{\widetilde \epsilon_{\bm P} -  \gamma_{\bm q}} \rp = 0
\ee
is satisfied.

\subsection{Square lattice}

Before treating specific cases, let us review the symmetries of the solution and the constraints imposed on it. We note that eq.~\eqref{eq:psi}  has a symmetry under simultaneous change $(\bm r, \bm P) \rightarrow (- \bm r, - \bm P)$ (global inversion symmetry). Another condition $\alpha(R,R) = 0$ translates into $\psi_{\bm P}(0) = 0$. Thus, we can search for solutions that are globally inversion symmetric and asymmetric, i.e. $ \psi^\pm_{\bm P }( \bm r) =  \psi_{\bm P }( \bm r) \pm \psi_{-\bm P }(- \bm r)$, respectively, such that $\psi^\pm_{-\bm P}(- \bm r) = \pm \psi^\pm_{\bm P }( \bm r)$.

In Fourier space, similarly, $\varphi^\pm_{-\bm P}(- \bm k) = \pm \varphi^\pm_{\bm P }( \bm k)$. This especially simplifies at $\bm P = 0$ and at $\bm P = (\pi, \pi)$, where the function becomes even or odd of its argument. 

\subsubsection{$\bm P = 0$}

We search for odd solutions, and note that there is only one irrep of   $C_{4v}$ group that is odd under in-plane inversion, which is two-dimensional. Denoting $\psi_1 = \psi(\bm t_1)$ and $\psi_2 = \psi(\bm t_2)$, we can rewrite \eqref{eq:lambda} as 
\be
\psi_1 \sin k_x +\psi_2 \sin k_y  =  \sum _{\bm q} \frac{ \gamma_{\bm k + \bm q} \lp \psi_1 \sin q_x +\psi_2 \sin q_y \rp }{\widetilde \epsilon_{\bm 0} -  \gamma_{\bm q}} = \sum _{\bm q} \frac{2 t \lp \cos (k_x + q_x) + \cos (k_y + q_y) \rp  \lp \psi_1 \sin q_x +\psi_2 \sin q_y \rp }{\widetilde \epsilon_{\bm 0} -  \gamma_{\bm q}}
\ee
where we used $\gamma_{\bm q} = 2 t (\cos q_x + \cos q_y)$. We expand the cosines of sums into products and some of the integrals are trivially zero due to symmetry constraints, and therefore we obtain 
\be
\lp \psi_1 \sin k_x   +\psi_2 \sin k_y  \rp \lp 1 + \sum_{\bm q} \frac{2 t \sin^2 q_x}{\widetilde \epsilon_{\bm 0} -  \gamma_{\bm q}} \rp =  0
\ee
where we used the symmetry of the integral under $x \Leftrightarrow y$. The term in the second brackets present self-consistency condition; the integrand is negative and maximum near BZ corners ($\bm q = (\pi, \pi)$), where its value is finite because the numerator vanishes quadratically. We find that the self-consistency equation is not satisfied for any $\epsilon_{\bm 0}$ below the band bottom (the integrals achieves its minimum value $\approx -0.36$ exactly at the band bottom energy), and therefore, we conclude that on a square lattice, there is no bound state at $\bm P = 0$.

\subsection{Triangular lattice}

As before, we consider the solutions with $\varphi^\pm_{-\bm P}(- \bm k) = \pm \varphi^\pm_{\bm P }( \bm k)$. This especially simplifies at $\bm P = 0$  the wavefunction becomes even or odd of its argument.

\subsubsection{$\bm P = 0$}

We search for odd solutions; there are two irreps (one-dimensional $\Gamma_3$ and two-dimensional $\Gamma_6$) of $C_{3v}$ group that is odd under in-plane inversion, which yield non-trivial function $\lambda(\bm k)$.

For $\Gamma_3$, we can take $\lambda(\bm k) = \psi_1 \lp \sin \bm k \cdot \bm t_1 + \sin \bm k \cdot \bm t_2 + \sin \bm k \cdot \bm t_3 \rp$, where $\bm t_1 = (1,0)$, $\bm t_2 = \lp-\frac{1}{2},\frac{\sqrt{3}}{2}\rp$, and $\bm t_3 = \lp-\frac{1}{2},-\frac{\sqrt{3}}{2}\rp$. Then eq.~\eqref{eq:lambda} turns into the self-consistency equation
\be
\sin \bm k \cdot \bm t_1 + \sin \bm k \cdot \bm t_2 + \sin \bm k \cdot \bm t_3 = \sum _{\bm q} \frac{2 t (\cos (\bm k + \bm q) \cdot \bm t_1 + \cos (\bm k + \bm q) \cdot \bm t_2 + \cos (\bm k + \bm q) \cdot \bm t_3) \lp \sin \bm q \cdot \bm t_1 + \sin \bm q \cdot \bm t_2 + \sin \bm q \cdot \bm t_3 \rp}{\widetilde \epsilon_{\bm 0} -  \gamma_{\bm q}}
\ee
where we used $\gamma_{\bm q} = 2 t (\cos \bm q \cdot \bm t_1 + \cos \bm q \cdot \bm t_2 + \cos \bm q \cdot \bm t_3)$. We notice that the terms in the numerator   that lead to nonvanishing integrals are  $-\sin \bm k \cdot \bm t_1 \sin \bm q  \cdot \bm t_1 \lp \sin \bm q \cdot \bm t_1 + \sin \bm q \cdot \bm t_2 + \sin \bm q \cdot \bm t_3\rp $ together with its cyclic permutation. Therefore, the self-consistency condition can be simplified down to
\be
1 + \sum_{\bm q} \frac{2 t \sin \bm q  \cdot \bm t_1 \lp \sin \bm q \cdot \bm t_1 + \sin \bm q \cdot \bm t_2 + \sin \bm q \cdot \bm t_3\rp }{\widetilde \epsilon_{\bm 0} -  \gamma_{\bm q}} = 0
\ee
As the value of $\epsilon_{\bm 0}$ approaches the band bottom ($-3 t$), the integrand diverges near the band minima, $K$ and $K'$. Therefore, the self-consistency condition can be easily saturated, which we find numerically to occur at $\epsilon_{\bm 0} \approx - 3.55 t$. Thus, we have found \emph{a bound state between a hole and a spin flip on a triangular lattice } which we call spin polaron, with binding energy $E_b = 0.42 t$, which is, remarkably, commensurate with $t$.

For two-dimensional representation $\Gamma_6$, the general form of $\lambda_{\bm 0}(\bm k)$ is $\widetilde \psi_1 \lp 2  \sin \bm k \cdot \bm t_1 -  \sin \bm k \cdot \bm t_2 - \sin \bm k \cdot \bm t_3 \rp + \widetilde \psi_2 \lp   \sin \bm k \cdot \bm t_2 - \sin \bm k \cdot \bm t_3 \rp$ for some independent constants $\widetilde \psi_{1,2}$ to be determined self-consistently. We plug this ansantz in, perform similar simplifications as before, and find the following self-consistency equation:
\be \label{eq:bound_state}
1 + \sum_{\bm q} \frac{2 t \sin \bm q  \cdot \bm t_2 \lp \sin \bm q \cdot \bm t_2 - \sin \bm q \cdot \bm t_3  \rp }{\widetilde \epsilon_{\bm 0} -  \gamma_{\bm q}} = 0
\ee
The integration-even part of the integrand is finite at $K,K'$ points because the numerator vanishes there, and the integral is bound from below by approx. $-0.5$. Thus, as we find, there is no bound state solution. Thus, the only solution that we find is for $\Gamma_3$ representation.

\begin{figure}[h] 
	\includegraphics[width= 0.5\columnwidth]{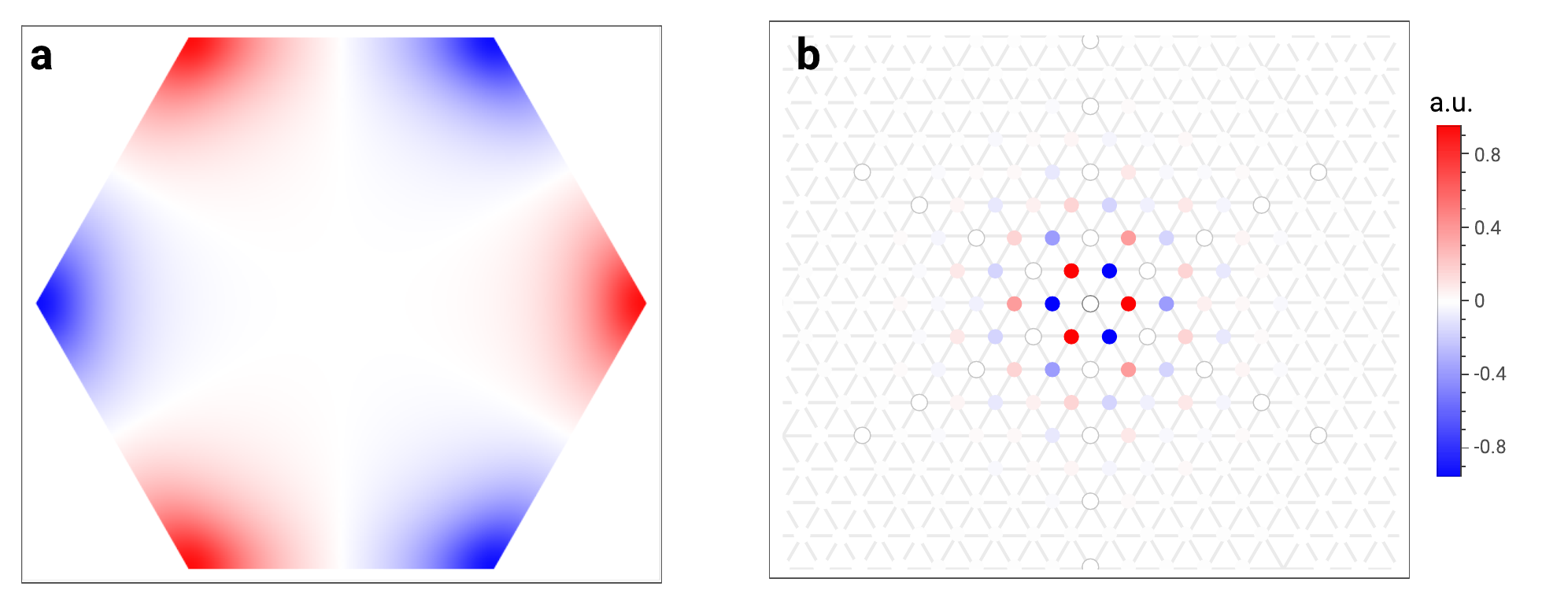}
	\caption{ (a) The $\bm k$-space and (b) the real-space plots of the wavefunction \eqref{master_sol} for the bound state found from \eqref{eq:bound_state}. This bound state is at $\bm P = 0$, realizes the symmetry of $\Gamma_3$ irrep of $D_6$ group and has a large binding energy $E_b \approx 0.42 t$, which explains its small localization length.
	} \label{fig:wavefunction}
\end{figure}

\begin{figure}[h] 
	\includegraphics[width= 0.35\columnwidth]{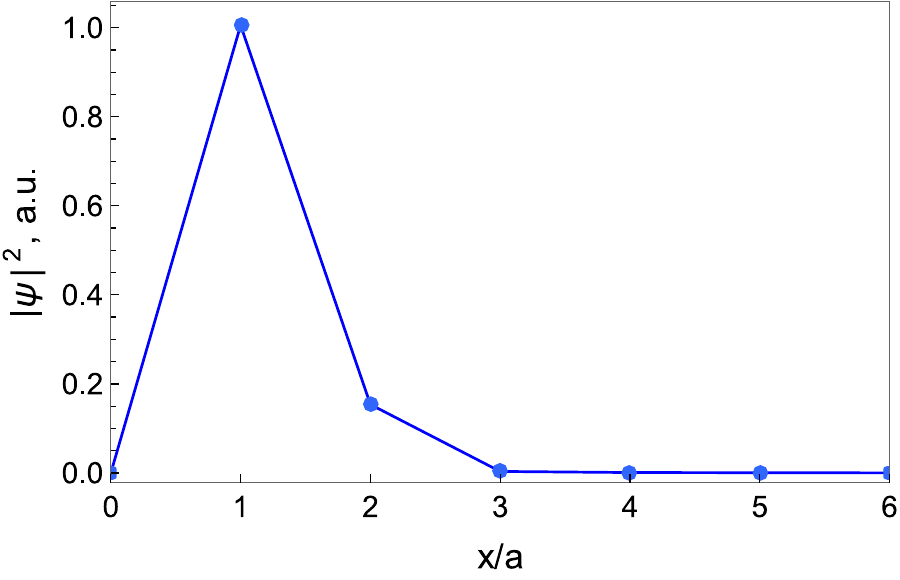}
	\caption{ Absolute value of the wavefunction from the previous plot along $x$-axis, illustrating its quick decay on the lengthscale of approximately one lattice spacing $\xi_{loc}\sim a$.
	} \label{fig:wavefunction}
\end{figure}

\subsubsection{Dispersion $E(\bm P)$}

In order to fins the dispersion of the bound state, we solve the problem numerically. For this, we recast the equation on the wavefunction:


%
\be
  \sum_n \psi_{\bm P} (-\bm r)  \lp e^{i \bm P \cdot \bm t_n}  \delta (\bm r - \bm t_n) + e^{-i \bm P \cdot  \bm t_n } \delta (\bm r + \bm t_n) \rp  +  \sum_n \lp  \psi_{\bm P} (\bm r + \bm t_n) + \psi_{\bm P} (\bm r - \bm t_n)\rp = \widetilde \epsilon_{\bm P} \psi_{\bm P} (\bm r)
\ee
\be \label{eq:num_sys}
\sum_{\bm r'}  \sum_n  \left [ \lp e^{i \bm P \cdot \bm t_n}  \delta (\bm r - \bm t_n) + e^{-i \bm P \cdot  \bm t_n } \delta (\bm r + \bm t_n) \rp \delta(\bm r' + \bm r)  +  \delta (\bm r' - \bm r + \bm t_n) + \delta (\bm r' - \bm r - \bm t_n)\right ]  \psi_{\bm P} (\bm r')  = \widetilde \epsilon_{\bm P} \psi_{\bm P} (\bm r)
\ee
We solve this equation numerically for a hexagon-shaped lattice with diameter up to 25 sites and periodic boundary condition to find the dispersion $E(\bm P)$. We also check numerically that for electron doping for triangular lattice, there is no bound state, and check that there is no bound state in the case of square lattice.     

\begin{figure}[h] 
	\includegraphics[width= 0.26\columnwidth]{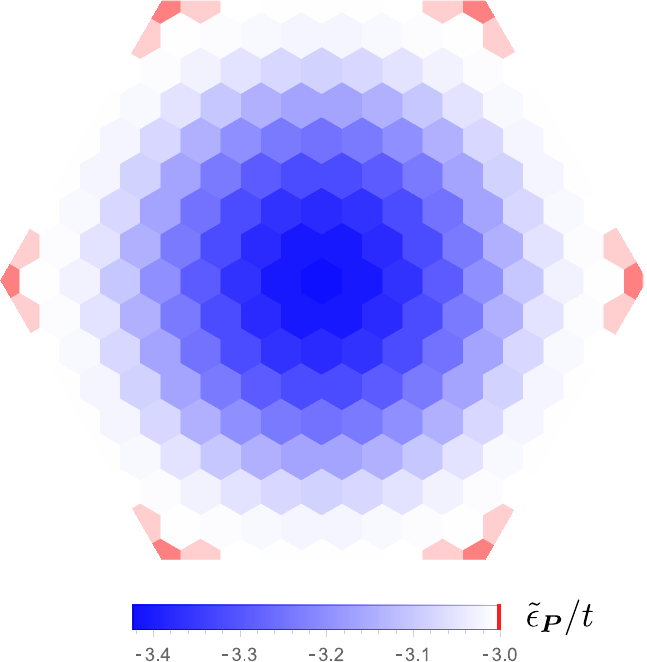}
	\caption{ Color map of the bound state dispersion $\widetilde{\epsilon}_{\bm P} $   in the Brillouin zone for $\bm P$.
	} \label{fig:numerics1_SM}
\end{figure}

\begin{figure}[h] 
	\includegraphics[width= 0.5\columnwidth]{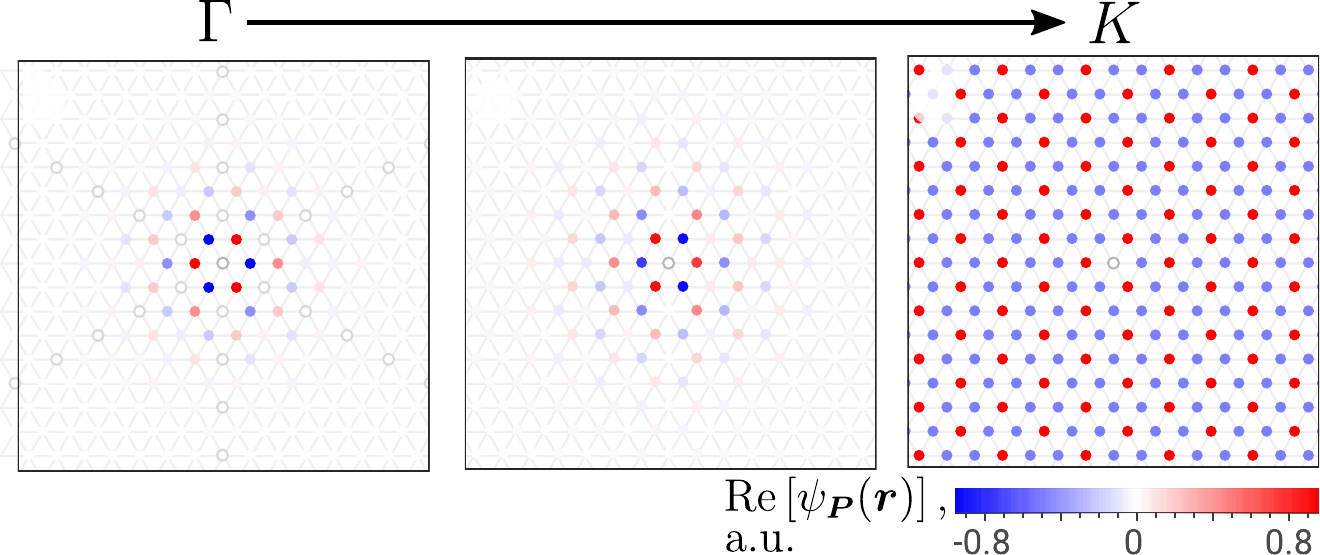}
	\caption{ The real-space wavefunction of the spin polaron in relative coordinates for several values of the center-of-mass momentum, $\bm P = \Gamma, (K-\Gamma)/2, K$.
	} \label{fig:wavefunction_multiple}
\end{figure}

The results of the numerical tight-binding calculations are shown in Fig.~\ref{fig:numerics1_SM}. We find that the spin polaron has a mass $m_{p} \approx 13 m_h$ near the $\Gamma$-point, where $m_h = \frac{2}{3 t a^2}$ is the mass of a hole. The wavefunction at finite momenta is shown in Fig.~\ref{fig:wavefunction_multiple}. The bound state exists everywhere except at the points $K,K'$. Upon approaching these points, the localization radius of the wavefunction increases.

\section{Spin polaron at finite $t/U$}

\subsection{Effective Hamiltonian}

In this part, we describe the tight-binding equation  solution for the spin polaron to the next leading order in $t/U$. First, we derive the effective Hamiltonian using the usual Hubbard-Stratonovich transformation. 
The hopping term in the Hamiltonian \eqref{H_SM} can be represented as a sum of three terms:
\be
H_t = T_0 + T_1 + T_{-1}
\ee
where the subscript denotes the change of the number of double occupied states in the corresponding hopping process. As a cartoon, all the possibilities can be expressed as the following cartoon:

\begin{figure}[h] 
	\includegraphics[width= 0.3\columnwidth]{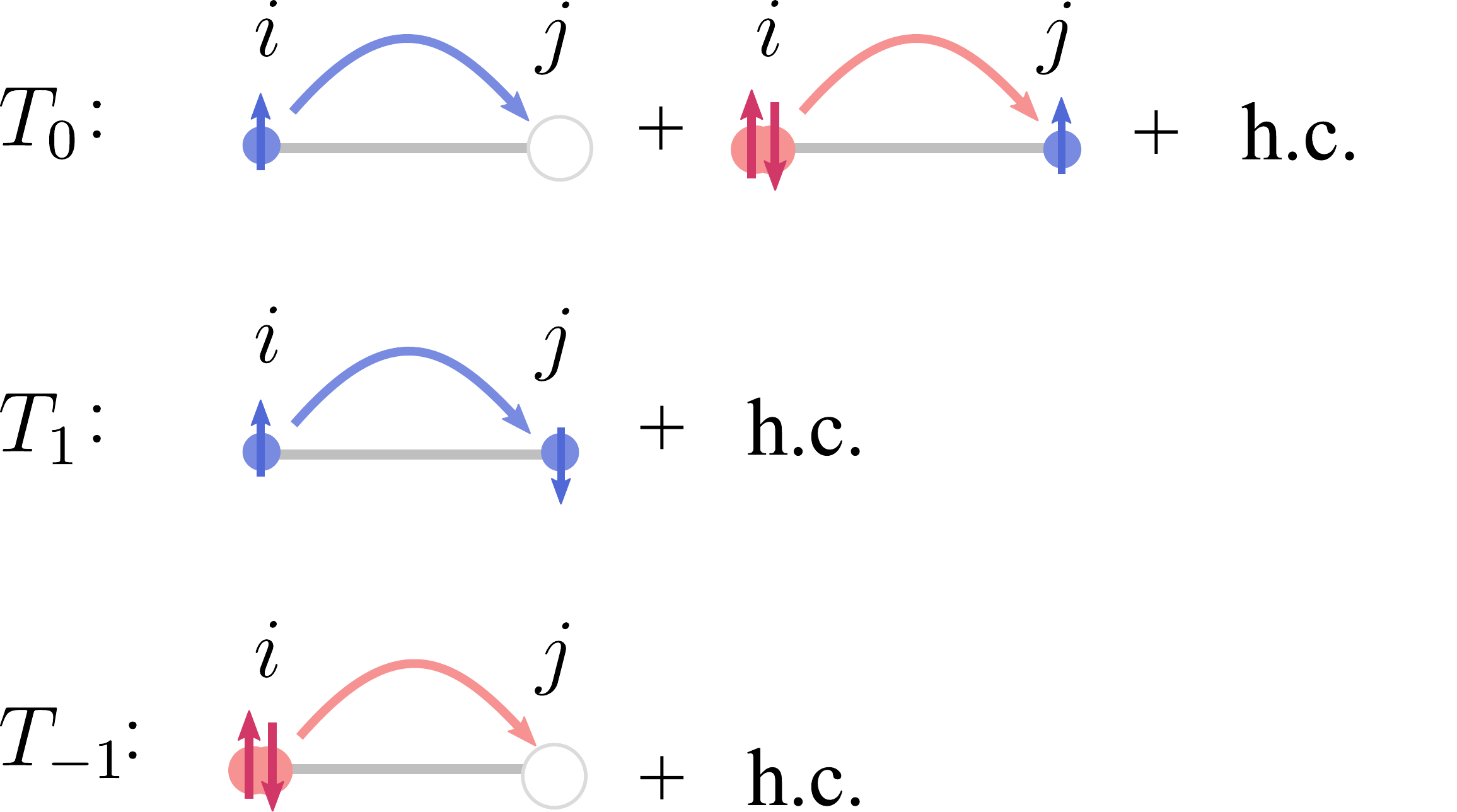}
	\caption{ 
	} \label{fig3}
\end{figure}

And can be written as an expression
\be
\begin{split}
T_0 &= -t\sum_{\langle i,j \rangle, \sigma}  c^\dagger_{j \sigma} c_{i \sigma} \lp  P_{j,\overline{\sigma}} P_{i,\overline{\sigma}} +  n_{j,\overline{\sigma}} n_{i,\overline{\sigma}} \rp + h.c.  \\
T_1 &= -t\sum_{\langle i,j \rangle}  c^\dagger_{j \sigma} c_{i \sigma} \  n_{j,\overline{\sigma}} P_{i,\overline{\sigma}}  + h.c. \\
T_{-1} &= -t\sum_{\langle i,j \rangle}  c^\dagger_{j \sigma} c_{i \sigma}   P_{j,\overline{\sigma}} n_{i,\overline{\sigma}} + h.c.
\end{split}
\ee
where $P_{i,\sigma} = 1 - n_{i,\sigma}$. 

Because the hopping commutes with the magnetic field terms, the commutator relations between the $T$-operators and $H_0$ are $[H_0,T_m] = m U T_m$. Therefore, we perform the transformation $H' = e^{i S} H e^{-i S}$ with $i S = \frac{T_1 - T_{-1}}{U}$, expand to the second power and obtain
\be
H' = H_0 + T_0 + \frac{[T_1,T_0] + [T_1,T_{-1}] +  [T_0,T_{-1}]}{U} + \mathcal{O}\lp \frac{t^3}{U^2} \rp
\ee
Because weassume  $U \gg t$, we can project the Hamiltonian on the subspace of the ground state and low-lying excitations, which corresponds to minimum double occupancy at given filling and total spin. Up to quadratic terms in $t$ this yields
\be
H_{eff} = H_0 + T_0  -\frac{T_{-1}T_1}{U}
\ee
For doped holes, the second term in $T_0$ in Fig. \ref{fig3} is absent.

We can further rewrite the Hamiltonian for the case of filling $\nu < \frac{1}{2}$ ($\delta n <0$). Projecting onto the manifold without double occupancies:
\be
\begin{split}
H_{eff} &=    h \sum_i \lp n_{i \uparrow} - n_{i \downarrow}\rp -t\sum_{\langle i,j \rangle, \sigma}  \lp  c^\dagger_{j \sigma} c_{i \sigma}  P_{j,\overline{\sigma}} P_{i,\overline{\sigma}}   + h.c. \rp  -\\
&- \frac{t^2}{U} \lp \sum_{\langle i,j \rangle,\sigma}  c^\dagger_{j \sigma} c_{i \sigma}   P_{j,\overline{\sigma}} n_{i,\overline{\sigma}} + h.c. \rp 
\lp \sum_{\langle i',j' \rangle, \lambda}  c^\dagger_{j' \lambda} c_{i' \lambda} \  n_{j',\overline{\lambda}} P_{i',\overline{\lambda}}  + h.c.\rp = \\
&=h \sum_i \lp n_{i \uparrow} - n_{i \downarrow}\rp -t\sum_{\langle i,j \rangle, \sigma}  \lp  c^\dagger_{j \sigma} c_{i \sigma}  P_{j,\overline{\sigma}} P_{i,\overline{\sigma}}   + h.c. \rp 
-\frac{t^2}{U} \sum_{\langle sj \rangle, \langle j i \rangle, \sigma} \lp c^\dagger_{s \sigma} c_{i \sigma} P_{j\sigma}n_{j \overline{\sigma}}P_{s\overline{\sigma}}P_{i\overline{\sigma}} + h.c. \rp - \\
&- \frac{t^2}{U} \sum_{\langle sj \rangle, \langle j i \rangle, \sigma} \lp c^\dagger_{s \overline{\sigma}}  c_{j \overline{\sigma}} c^\dagger_{j \sigma} c_{i \sigma} P_{s\sigma}P_{i\overline{\sigma}} + h.c. \rp
\end{split}
\ee
We can split the last two terms to the cases when $s = i$ and $s \neq i$:
\be \label{eq:Heff_gen_holes}
\begin{split}
H_{eff} &=   h \sum_i \lp n_{i \uparrow} - n_{i \downarrow}\rp -t\sum_{\langle i,j \rangle, \sigma}  \lp  c^\dagger_{j \sigma} c_{i \sigma}  P_{j,\overline{\sigma}} P_{i,\overline{\sigma}}   + h.c. \rp 
-\\
&-\frac{t^2}{U} \sum_{\langle ij \rangle, \sigma}  n_{i \sigma} n_{j \overline{\sigma}}   + \frac{t^2}{U} \sum_{\langle ij \rangle, \sigma} \lp c^\dagger_{i \overline{\sigma}} c_{i \sigma}  c^\dagger_{j \sigma} c_{j \overline{\sigma}}   + h.c. \rp
\\
&-\frac{t^2}{U} \sum_{\left \{  i,j,s \right \},\sigma} \lp c^\dagger_{s \sigma} c_{i \sigma} P_{j\sigma}n_{j \overline{\sigma}}P_{s\overline{\sigma}}P_{i\overline{\sigma}} + h.c. \rp   
- \frac{t^2}{U} \sum_{\left \{  i,j,s \right \},\sigma} \lp c^\dagger_{s \overline{\sigma}}  c_{j \overline{\sigma}} c^\dagger_{j \sigma} c_{i \sigma} P_{s\sigma}P_{i\overline{\sigma}} + h.c. \rp
\end{split} 
\ee
The second line turns into an effective nearest-neigbor interaction and the spin part of the Hamiltonian, as we show below. The last line corresponds to a density-mediated hopping and a simultaneous hopping of a correlated hole and a spin flip. In the literature they are called either $t-J-3s$ model or $t-J-t3$ model. These terms are referred to as three-site terms, pair hopping or conditional hopping terms. In the absence of magnetic field they have been shown to facilitate superconducting phase and push away phase separation phase. 

The terms in the last line only occur when a hole neighbors with a spin flip in an otherwise ferromagnetic (or ferrimagnetic) environment that we want to consider. The "three-site" terms are illustrated in the cartoon below.

\begin{figure}[h] 
	\includegraphics[width= 0.6\columnwidth]{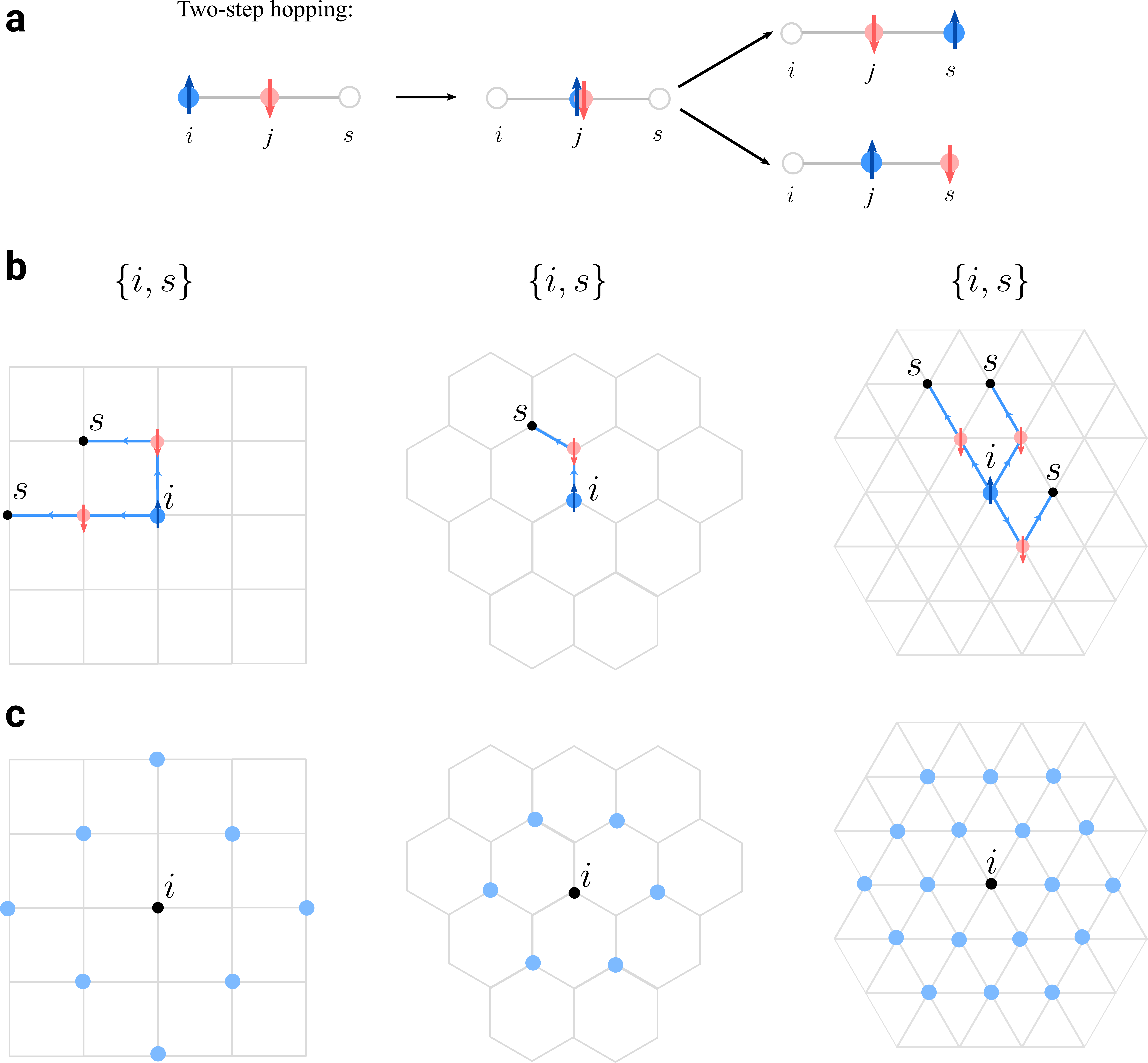}
	\caption{ (a) Two options for the correlating hopping  terms from site $i$ to site $s$. (b) 
	The types of the paths for three-site terms starting from site $i$.
	(c) The set of final sites as seen from site $i$ for three-site processes.
	} \label{fig4}
\end{figure}

Notice that on frustrated lattices the sites to which the hole can hop via three-site process include  nearest-neighbors. On bipartite lattice such hopping can occur only within the same sublattice.

\subsection{Equation for the bound state}

\begin{figure}[t] 
	\includegraphics[width= \columnwidth]{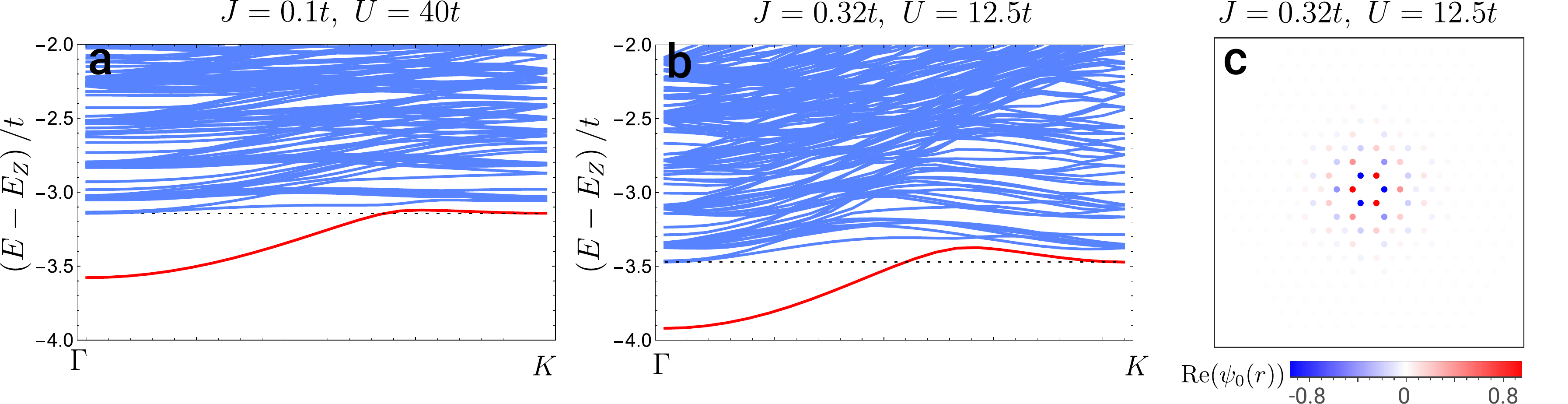}
	\caption{ (a,b) The energy spectrum of the Hubbard model at $n=1$ doped with one hole as a function of the center of mass momentum $\bm P$ along $\Gamma-K$ direction at finite values of $U$. The spectrum was obtained by solving the tight-binding equation \eqref{eq:psi_SM_new}.  (c) The real space plots of the wavefunction at momentum  $\bm P = \Gamma$. 
	} \label{fig:SM_1}
\end{figure}

We use the same ansatz given in eq.~\eqref{ansatz_SM}, and derive the new SE for the wavefunction:
\be
\begin{split}
&\frac{3}{2} h \alpha_{\ell s} + t \sum_n \alpha _{s \ell } (\delta(\ell-s - \bm t_n) + \delta(\ell-s + \bm t_n))+ t \sum_n (\alpha _{\ell + \bm t_n, s} + \alpha _{\ell- \bm t_n, s}) 
\\
&- \frac{t^2}{U} \lp 2 \alpha_{ls} - 2 \sum_{\pm \bm t_n}  \left [ \alpha_{ls} \delta (l - s - \bm t_n) + \alpha_{l, s+ \bm t_n}  \right ] + \right. 
\\
&+  \sum_{\pm \bm t_n , \pm \bm t_m, \neq \bm t_m}     \left[  -\alpha_{k + \bm t_m - \bm t_n} \delta(l-s-\bm t_n) +
\alpha_{s-\bm t_m + \bm t_n, s - \bm t_m } \delta(l - s - \bm t_n + \bm t_m)   \right.
\\
&+ \left.\left. \alpha_{l - \bm t_n + \bm t_m, s - \bm t_n} \delta(l - s - \bm t_n) - \alpha_{l+ \bm t_m - \bm t_n,l} \delta(l-s - \bm t_n + \bm t_m)
\right ]   \rp 
\\
&= E \alpha_{\ell s}
\end{split}
\ee
We again separate the coordinates into the one of center of mass and the relative one using $\alpha_{nm} = \alpha(\bm R + \bm r, \bm R)$, where $\bm R = \bm r_m$ and $\bm r = \bm r_n - \bm r_m$. We then  use $\alpha( \bm R + \bm r, \bm R)  =  \sum_{\bm P}  \psi_{\bm P}(\bm r)  e^{i \bm P \cdot \bm R}$ in order to find a tight-binding equation for the solution with a specific center of mass momentum:
\be \label{eq:psi_SM_new} 
\begin{split}
&\lp E - \frac{3}{2} h + 2 \frac{t^2}{U} \rp  \psi_{\bm P} (\bm r) - t \sum_{\pm t_n} \lp1 + 2 \lp\frac{t}{U} \rp e^{-i \bm P \bm t_n}\rp   \psi_{\bm P} (\bm r + \bm t_n)  
\\
&= t \sum_{\pm t_n}  \psi_{\bm P} (-\bm r) e^{i \bm P \cdot \bm t_n} \delta (\bm r - \bm t_n) + 2 \frac{t^2}{U} \sum_{\pm t_n} \psi_{\bm P} (\bm r) e^{ \bm P \cdot \bm t_n} \delta (\bm r - \bm t_n) 
+
 \frac{t^2}{U} \sum_{\pm \bm t_n, \pm \bm t_m; \ \bm t_n \neq \bm t_m} \psi_{\bm P} (\bm t_n) \delta(\bm r - \bm t_m) 
\\
&- \frac{t^2}{U} \sum_{\pm \bm t_n, \pm \bm t_m; \ \bm t_n \neq \bm t_m} \lp    
\psi_{\bm P} (\bm t_n) \delta(r-\bm t_m + \bm t_n) e^{-i \bm P \bm t_n}
+
 \psi_{\bm P} (\bm t_n - \bm t_m) \delta(\bm r + \bm t_n) e^{-i \bm P \bm t_n}
 -
 \psi_{\bm P} (-r) \delta(r-\bm t_n + \bm t_m) e^{i \bm P (\bm t_m - \bm t_n)}
\rp
\end{split}
\ee
The solution to these equations for momentum $\bm P$ along $\Gamma-K$ direction at finite values of $U$ is shown in Fig.~\ref{fig:SM_1}. The bottom of the continuum is shown by the black dashes line, and the gap from the bottom of the bound state dispersion to the continuum occurs at $\Gamma$-point and gives the binding energy of the spin polaron. The spectrum changes more as the $U$ is decreased further and eventually develops two band minima for the spin polaron dispersion.  The wavefunction is plotted in panel (c).

\begin{figure}[h] 
	\includegraphics[width= 0.43\columnwidth]{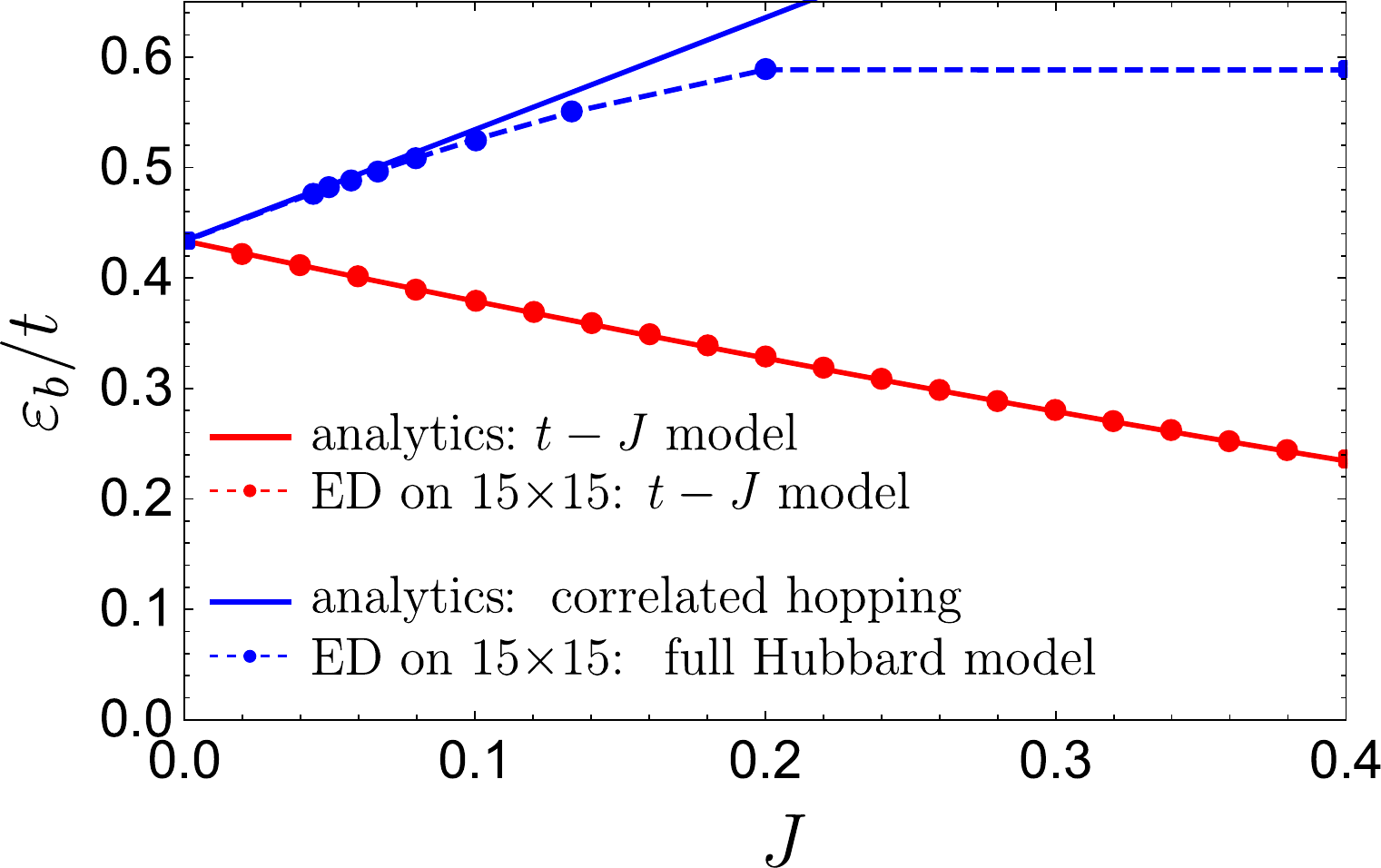}
	\caption{The dependence of the binding energy of the spin polaron as a function of $J = 4 t^2/U$ showing an increase of the binding energy as a function of $J$ for the realistic description and a decrease for the $t-J$ model. The results of the exact diagonalization for the full Hubbard model and the $t-J$ model are shown by blue and red points, correspondingly. The  results obtained from the analytical approach introduced in this section for the effective Hamiltonian at order $t^2/U$ and for the $t-J$ model are shown by blue and red solid lines.
	} \label{fig:SM_2}
\end{figure}

For the binding energy, we perform ED calculation within Hubbard model on a $15\times 15$ lattice and obtain the behavior shown by the dashed line in Fig.~\ref{fig:SM_2}. The solution obtained from the analytical approach above at the order $t^2/U$ is shown by the solid blue line and agrees with the full ED calculation as $t/U$ is decreased. Surprisingly, one observes that as $t/U$ ($J$) increases, the binding energy of the spin polaron also increases. 

In order to understand this behavior better, we compare these results with a calculation within $t-J$ model, which captures  only the spin exchange at the level $t^2/U$ (the result for the $t - t_3 -J$ model appeared earlier in the literature \cite{PhysRevB.97.140507}). For this model, as expected, the ED and the analytical solution (which neglects the correlated hopping contributions) results coincide. For the $t-J$ model, the binding energy decreases, because now the spin loses some of the kinetic energy by binding to the hole, and eventually, vanishes around $J \approx 11 t$. 
Therefore, it further supports the conclusion that the increase of the binding energy of the spin polaron as $t/U$ increases occurs due to the effect of the correlated hopping; it introduces additional processes that allow the bound state to further gain mobility due to the proximity of the spin to the hole. Because the formation of the spin polaron is a microscopic process, it is natural that excluding the correlated hopping terms at $t^2/U$ order is not justified anymore.

\section{Metallic magnetism at finite doping}
At finite doping density, we study the metallic magnetism and critical magnetic field for the fully polarized state in the infinite-U limit for triangular lattice. Without double occupancy, the infinite-U Hubbard model is reduced to the $t-J$ model with $J=0$ as shown in Eq. S7. We further consider the case of finite $J=0.1t$ as following:
\begin{equation}
\hat{H}=-t \sum_{\langle i j\rangle, \sigma}\left(c_{i \sigma}^{\dagger} c_{j \sigma}+\text { h.c. }\right)+ J \sum_{\langle i j\rangle}\left(\vec{S}_{i} \cdot \vec{S}_{j}-\frac{n_{i} n_{j}}{4}\right)
\end{equation}


Exact diagonalization is employed to calculate the magnetic ground state with finite doping density up to system size $3 \times 18$ for doping density less than $1/3$. To reduce the dimension of Hilbert space, the full Hamiltonian is divided into momentum and spin sectors by translation symmetry and spin conserving. Comparing the energy difference between different spin sectors, we can extract the critical magnetic field for fully polarized state. For the case of one hole doping and one spin flip, the simulated system size goes up to $24\times 24$. We cross check the results with QuSpin program, especially for infinite U limit \cite{weinberg2017quspin}. 

\begin{figure}[h] 
	\includegraphics[width= 0.5\columnwidth]{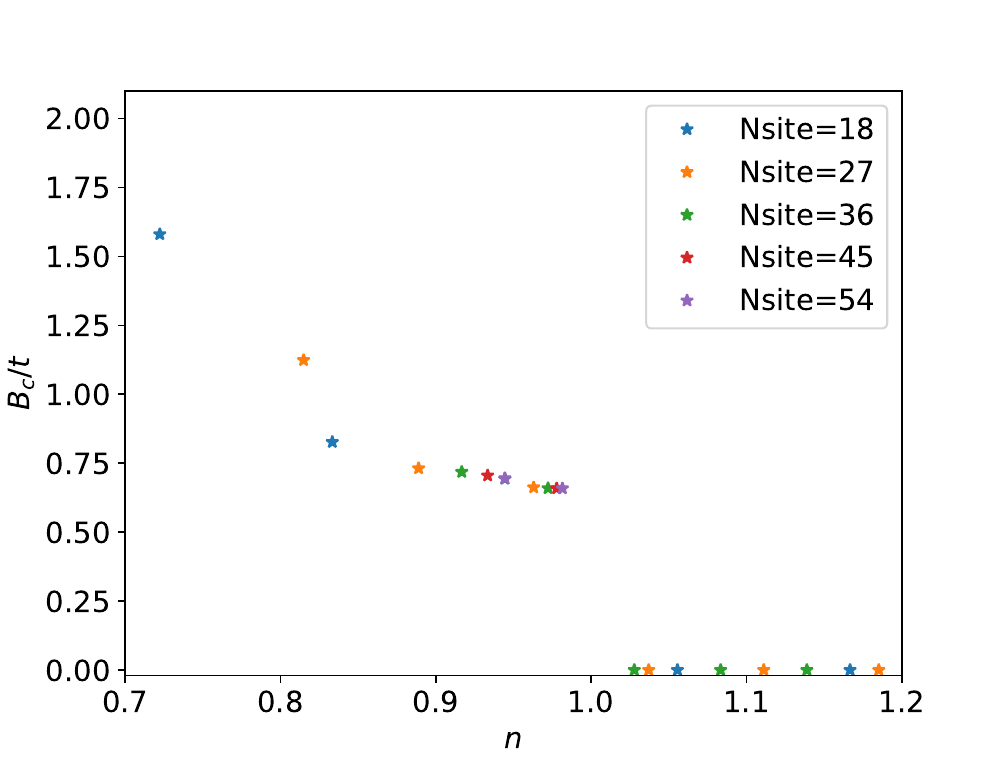}
	\caption{ The saturation field $h_s$ at different filling densities $n$ calculated by exact diagonalization on 3$\times$6, 3$\times$9, 3$\times$12, 3$\times$15, 3$\times$18 geometries in the limit $U = \infty$. For all calculations, we only present the data points for odd number of electrons or holes.
	} \label{fig:ED_suppl}
\end{figure}

At the hole doping side, the ground state is antiferromagnetic, we only consider the case of $0,1,2$ spin slip for the extrapolation of saturation field. For $(L_x,L_y) =(3,6)$ and $(3,9)$ with two and three holes, the two spin flip state is lower in energy than one spin flip when decreasing the magnetic field below $h_s$. In Fig. \ref{fig:ED}, we plot the critical magnetic field vs. doping density for system size $(L_x,L_y) =(3,6), (3,9), (3,12), (3,15), (3,18)$ with odd number of electrons and holes from n=1. It is found that critical field for fully polarized state is increasing with the hole density up to $1/3$ doping. For system size $(18,18)$, the saturation field for one hole doping is $0.423t$, approaching the analytical value $h^*=0.42t$.

\end{document}